\documentclass[a4paper,10pt,aps,prd,twocolumn]{revtex4}
\usepackage[colorlinks,linkcolor={red},citecolor={blue}]{hyperref}
\usepackage{graphicx}
\usepackage{amsmath}
\usepackage{amssymb}
\usepackage{amsfonts}
\usepackage{bm}
\usepackage{braket}
\usepackage{color}

\def\be{\begin{align}}
\def\ee{\end{align}}
\def\bea{\begin{eqnarray}}
\def\eea{\end{eqnarray}}

\newcommand{\mc}[1]{\mathcal{#1}}
\newcommand{\f}[2]{\frac{#1}{#2}}

\begin{document}
	
\title{Cosmology of the non-local Proca theory}
\author{Zahra Haghani}
\email{z.haghani@du.ac.ir}
\affiliation{School of Physics, Damghan University, Damghan, 
	41167-36716, Iran.}

\begin{abstract}
	In this paper, we will consider cosmological implications of the Maxwell theory coupled to a non-local $U(1)$-symmetric term. It is well-known that the theory in flat space time, reduces to the Proca theory. However, it will be shown that in curved space time the resulting theory will differ from the coupled Einstein-Proca system. The cosmological perturbations on top of the de Sitter space-time is also considered and the dynamics of separate modes will be investigated in details. Anisotropic cosmology of the model is also investigated and we will show that the behavior of the universe at late time satisfies the observational data and the model predicts an isotropic universe.
\end{abstract}


\maketitle

\section{Introduction}

The late-time accelerated expansion of our universe is demonstrated by different observational data \cite{acce}. The late-time acceleration occurs at very low energy scale $\sim 10^{-4}eV$, so to investigate this phenomenon one can consider a modification to GR as an effective field theory of gravity. There are many attempts in the literature to describe the observational data. The main idea is adding new degrees of freedom to general relativity (GR) in a consistent way so that the resulting theory does not contain instabilities and also satisfy current observational data.  
Adding new degrees of freedom can be done in different approaches. The first approach is to introduce some matter sources with a large negative pressure. This branch is the well-known as dark energy branch. There are many types of such sources in the literature such as scalar fields, vector fields and etc \cite{DEmodels}. The $\Lambda$CDM model is the simplest and most successful model to describe current acceleration of the universe and this model is in a good agreement with observational data. However this model has some theoretical and phenomenological problems, motivated cosmologists to consider more complicated cases containing dynamical extra degrees of freedom \cite{lamp}.
 The second approach is to modify the gravity sector in Riemannian geometry at  large distances such as $f(R)$, scalar-tensor theories \cite{modg}. Also, one can go further and make a theory with a non-minimal couplings between matter and geometry such as  $f(R,T)$ \cite{fRT}, $f(R,T,R_{\mu\nu}T^{\mu\nu})$ \cite{FRTR}, $f(R,L_m)$, etc \cite{matt}. Cosmological implications of such theories are well-known and investigated extensively in the literature. Other possibilities in this branch is to consider non-Reimannian geometries such as Weyl-Dirac theory \cite{weyl}, Cartan geometry \cite{cartan} and their generalizations \cite{genwc}.
 
In this work we will investigate a special class of a vector-tensor theory of gravity which contains nonlocal self-interactions of the vector field. The application of vector field theories in cosmological behavior of the universe is well-known and the consequences of such an extension is investigated vastly both for early times \cite{vecinf} and also for the late time cosmology \cite{veclate}.  More important possibilities for the vector field Lagrangian would include the Maxwell theory describing a massless vector field which has U(1) symmetry and the Proca theory for the massive vector field. 

The kinetic term of the Maxwell and also Proca theories $-1/4F_{\mu\nu}F^{\mu\nu}$ has a famous problem that it diverges at the point charge. There are some efforts in the literature to avoid this problem. Born and Infeld \cite{BI} have considered non-linear terms of the strength tensor of the form
$$\mathcal{L}\propto1-\sqrt{\det\left(\eta_{\mu\nu}+\beta F_{\mu\nu}\right)},$$
to avoid the point charge divergence of the electric charge. The Born-Infeld electrodynamics reduces to the Maxwell's theory in low energy limits. The cosmological implications of this kind of matter field has been vastly investigated \cite{BIcos}.

The other possibility for avoiding the aforementioned divergence is to add some higher derivative self-interaction terms to the theory. This was first done by Bopp and Podolsky \cite{BP} where their Lagrangian contains
\begin{align}
\mathcal{L}\propto-\f{1}{m^2}\big( \partial_\alpha F^{\mu\nu}\big)^2,
\end{align} 
where $m$ is a constant with mass dimension 1. After a field redefinitions it can be shown that the Bopp-Podolsky Lagrangian describes two independent dynamical fields corresponding to one massless and one massive vector field. Because of the higher derivative nature of the Bopp-Podolsky interaction term, the theory contains Ostrogradski instability which shows itself as a massive vector ghost. One should note that both generalizations of the Maxwell theory introduced above, respects U(1) symmetry. Cosmological implications of the Bopp-Podolsky theory and also its ghost instability on top of de Sitter expanding background has been investigated in \cite{BPcos}. 

In an effective action the origin of nonlocal terms can be seen in the one-loop approximation via the heat kernel method. Also one can consider the nonlocal contributions to the quantum effective action within the covariant perturbation expansion in the field strengths and spacetime curvatures \cite{barvinsky}.
From an independent viewpoint, nonlocal terms in classical action may come from integrating out one of the healthy degrees of freedom of the theory. In these cases the resulting  nonlocal action is free from pathologies.
Also there are special higher dimensional gravity theories, in which reduction to four dimensions gives an action including nonlocal terms, e.g. $\sqrt{-\Box}$ in DGP model \cite{DGP}. Nonlocal terms has been used to modify IR and UV limits of GR  and nonlocal cosmology is extensively investigated in the literature \cite{noncos}.

Recently, nonlocal terms have been extensively studied in massive field theories and specially massive gravity. In these cases, one can retain the gauge invariance of the theory by introducing some nonlocal  terms. Nonlocal terms for massive spin two fields was suggested in the context of degravitation \cite{degra}.
Also, nonlocal massive gravity was studied in \cite{Maggior}. In this paper, we will consider the effect of nonlocal terms in a vector field theory.
 
It is well-known that the massive vector theory can be written in a form that it is explicitly $U(1)$ symmetric. This can be done by the additional dynamical degree of freedom which behaves correctly under the $U(1)$ transformation. To illustrate this point further, let us consider a dynamical massive vector field in flat space-time described by the Proca action of the form
\begin{align}\label{action}
S=\int d^4x\left(-\f14F_{\mu\nu}F^{\mu\nu}-\f12m^2A_\mu A^\mu \right),
\end{align}
where  $F_{\mu\nu}=2\partial_{[\mu}A_{\nu]}$ is the field strength of the Proca field $A_\mu$ with dimension $M$ and $m^2$ is the mass squared. This theory has 3 degrees of freedom on flat space and breaks explicitly the $U(1)$ symmetry of the Maxwell's theory due to existence of the mass term. Now, let us Stueckelberg transform the field $A_\mu$ as
$$A_\mu\rightarrow A_\mu+\partial_\mu\phi,$$
where $\phi$ is a scalar field. The resulting action becomes
\begin{align}\label{action1}
S=\int d^4x\bigg(&-\f14F_{\mu\nu}F^{\mu\nu}-\f12m^2A_\mu A^\mu\nonumber\\&-m^2\partial_\mu\phi A^\mu-\f12 m^2\partial_\mu\phi\partial^\mu\phi \bigg),
\end{align}
which has an explicit $U(1)$ symmetry if both $A_\mu$ and $\phi$ fields transform as
\begin{align}
A_\mu&\rightarrow A_\mu+\partial_\mu\zeta,\nonumber\\
\phi&\rightarrow\phi-\zeta,
\end{align}
where $\zeta$ is an arbitrary function of coordinates.
One can proceed further and write the action \eqref{action1} in a form that it contains only the field $A_\mu$ and also has an explicit $U(1)$ symmetry. This can be done by obtaining the scalar field $\phi$ in terms of the vector field $A_\mu$ from the scalar equation of motion
$\Box\phi=-\partial_\mu A^\mu,$
with the result
\begin{align}
\phi=-\f{1}{\Box}(\partial_\mu A^\mu),
\end{align}
and substituting back to the action \eqref{action1}. The result is
\begin{align}\label{non}
S=\int d^4x\bigg(&-\f14F_{\mu\nu}F^{\mu\nu}+\f{m^2}{4}F_{\mu\nu}\f{1}{\Box}F^{\mu\nu}\bigg),
\end{align}
which we have performed integration by parts to simplify the action. One can see that the resulting action has a  $U(1)$ symmetry, but it becomes non-local.
As a result, the Proca theory in flat space is equivalent to non-local theory \eqref{non}.

The generalization of the Proca action has also attracted considerable attention in the literature \cite{gp}. In these theories the general kinetic term for a massive vector field beyond Maxwell term is considered which has 3 degrees of freedom. Generalization to the case of non-Abelian vector field is also considered in \cite{nonpro}.

The line of the paper is as follows. In the next section, we will introduce the model and obtain the field equations and also discuss on the local counterpart of the theory. In section \ref{iso} we will investigate the isotropic cosmology of the model and consider its cosmological perturbation on top of de Sitter solution. Section \ref{aniso} will be devoted to the anisotropic implications of the theory and in the last section we will conclude and discuss on some possible issues and future lines.

\section{The model}\label{sec2}
We will consider the effects of a Proca field in cosmology, but we want to make the theory $U(1)$ symmetric. This can be achieved by using the non-local counterpart of the Proca action. As a result, consider the following action
\begin{align}\label{ac}
S=\int d^4x\sqrt{-g}&\bigg(\kappa^2(R-2\Lambda)\nonumber\\&-\f14F_{\mu\nu}F^{\mu\nu}+\sum_n\alpha_n F_{\mu\nu}\f{1}{\Box^n}F^{\mu\nu}\bigg),
\end{align}
where we have generalized the non-local term to contain higher order derivatives and $\Lambda$ is the cosmological constant. Here, $\alpha_n$ is an arbitrary constant with dimension $M^{2n}$. 

In this paper we will consider the case $n=1$ for the sake of simplicity. The above theory with $n=1$, can also be considered as a non-local version of the so-called BP electrodynamics \cite{BP,BPcos,farsi}. The procedure of obtaining \eqref{non} from \eqref{action},  highly depends on the geometry of space-time. In fact partial derivatives should commute to make these two actions equal. In curved space-times the non-local version of the Proca field \eqref{ac} is different from the Einstein-Proca system by some non-linear curvature terms. One can easily show that the non-local part of the action \eqref{action} for $n=1$ can be written as
\begin{align}
\int d^4 x &\sqrt{-g}\,\alpha\,F_{\mu\nu}\f{1}{\Box}F^{\mu\nu}=\int d^4 x \sqrt{-g} \bigg[-2\alpha A^2\nonumber\\&+2\alpha A_\mu\Box^{-1}\bigg(R^{\mu\nu}A_{\nu}+\big(\f12\nabla_\nu R+R_{\alpha\nu}\nabla^{\alpha}\big)\Box^{-1} F^{\nu\mu}\nonumber\\&
~~~+\big(\nabla^\beta R^\mu_{~\alpha\beta\nu}+2R^{\mu}_{~\alpha\beta\nu}\nabla^{\beta}\big)\Box^{-1} F^{\alpha\nu}
\bigg)\bigg],
\end{align}
where we have used the condition $\nabla_{\alpha}A^{\alpha}=0$. The first term in the above expression is the mass term in the Proca action. All other terms contain the Riemann tensor and its contractions, which vanish on top of flat space-time.
 Another way to make this action local, is to rewrite the action \eqref{ac} with $n=1$ as 
\begin{align}\label{ac1}
S=\int d^4x \sqrt{-g}&\bigg(\kappa^2(R-2\Lambda)-\f14F_{\mu\nu}F^{\mu\nu}\nonumber\\&+\alpha F^{\mu\nu}\xi_{\mu\nu}+\lambda^{\mu\nu}(\Box\xi_{\mu\nu}-F_{\mu\nu})\bigg),
\end{align}
where $\xi_{\mu\nu}$ and $\lambda_{\mu\nu}$ are two covariant antisymmetric tensors.
 In the above action we have used the definition $\xi_{\mu\nu}=\Box^{-1}F_{\mu\nu}$ and implied this definition to the action through Lagrange multiplier $\lambda_{\mu\nu}$. So the tensor field $\xi_{\mu\nu}$ carries the degrees of freedom of the vector field $A_{\mu}$ and it should not have new degrees of freedom otherwise the action \eqref{ac1} would be different from  the non-local action \eqref{ac} with $n=1$.

Let us now obtain the equations of motion of the theory \eqref{ac1}. The equation of motion for the vector field $A_\mu$, the tensor field $\xi_{\mu\nu}$ and the Lagrange multiplier $\lambda_{\mu\nu}$ can be written respectively as
\begin{align}\label{eq1}
	\nabla^\beta(F_{\alpha\beta}-2\alpha\xi_{\alpha\beta}+2\lambda_{\alpha\beta})=0,
\end{align}
\begin{align}\label{eq2}
\alpha F^{\alpha\beta} +\Box\lambda^{\alpha \beta}=0,
\end{align}
and
\begin{align}\label{eq3}
\Box\xi_{\alpha \beta}=F_{\alpha\beta}.
\end{align}
The general  solution of field equation \eqref{eq3} related to the homogeneous equation  $\Box\xi_{\alpha \beta}=0$ would introduce new degrees of freedom to the theory which are independent of $A_\mu$. However as mentioned before these are spurious. So we should remind that we do not need the most general solution $\xi_{\alpha \beta}$  that satisfy equation \eqref{eq3}, but need the specific solution of the equation \eqref{eq3} with the condition that $\xi_{\mu\nu}=0$ as $F_{\mu\nu}=0$. In the quantum point of view there are no quanta associated to the auxiliary tensor field $\xi_{\alpha \beta}$ \cite{mag}.

Also it is worth mentioning that the theory has a $U(1)$ symmetry by shifting the vector field $A_\mu\rightarrow A_\mu+\partial_\mu\varphi$. This implies that the $A_\mu$ field equation satisfies a conservation equation of the form \eqref{eq1}. As, a result, we have a Noether current associated with the aforementioned symmetry. However the local Proca action does not have a gauge invariance.

The gravitational field equation of motion can be written as
\begin{widetext}
\begin{align}\label{eq4}
&\kappa^2( G_{\mu \nu}+ \Lambda \
g_{\mu \nu})- \frac{1}{2} F_{\mu}\text{}^{\alpha} F_{\nu \alpha}   + 2\alpha F_{(\mu}\text{}^{\alpha} \xi_{\nu ) \alpha} -  \tfrac{1}{2} g_{\mu \nu} \nabla_{\gamma}\xi_{\alpha \beta} \
\nabla^{\gamma}\lambda^{\alpha \beta} + \frac{1}{2} F^{\alpha \beta} \left(\f14 F_{\alpha \beta} -  \lambda_{\alpha \beta} -   \alpha 
\xi_{\alpha \beta}\right)g_{\mu \nu}
-   \xi_{(\mu}\text{}^{\alpha} \Box\lambda_{\nu )\alpha}\nonumber\\
& + \lambda_{(\mu}\text{}^{\alpha} \Box\xi_{\
	\nu) \alpha} +  \nabla_{(\mu}\lambda^{\alpha \beta} \nabla_{\nu)}\xi_{\alpha \beta} +\nabla^{\beta}\left[\lambda_{(\mu}^{~~\alpha}\nabla_{\nu)}\xi_{\alpha\beta}-\xi_{\alpha\beta}\nabla_{(\mu}\lambda_{\nu)}^{~~\alpha}+\xi_{(\mu}^{~~\alpha}\nabla_{\nu)}\lambda_{\alpha\beta}-\lambda_{\alpha\beta}\nabla_{(\mu}\xi_{\nu)}^{~~\alpha}\right] =0,
\end{align}
\end{widetext}
where parenthesis denotes symmetrization $A_{(\mu\nu)}=(A_{\mu\nu}+A_{\nu\mu})/2$.
 In the following the cosmological solution of this model is considered.

\section{Isotropic cosmology}\label{iso}
Let us now assume that the universe can be described by a flat FRW line element of the form
\begin{align}
ds^2=-dt^2+a^2(dx^2+dy^2+dz^2),
\end{align}
where $a$ is the scale factor. The homogeneity and isotropy of the FRW metric implies that the vector field $A_\mu$ should have a form
\begin{align}\label{aa}
A_{\mu}=(A_0(t),0,0,0).
\end{align}
This form is the only choice which preserve isotropy and homogeneity of the universe. The antisymmetric tensor fields $\lambda_{\mu\nu}$ and $\xi_{\mu\nu}$ has six degrees of freedom consists of a 3-vector corresponding to the (0i) components and a pseudo 3-vector field corresponding to (ij) components \cite{maggoire}. However, in the isotropic and homogeneous FRW universe these tensors should vanish, because otherwise there is a non-zero 3-vector field which breaks the isotropy of the space-time. So, in the FRW universe, we have $\lambda_{\mu\nu}=0=\xi_{\mu\nu}$, (for the detailed proof see the appendix \ref{app}). Also note that the form \eqref{aa}  for the vector field implies that the field strength $F_{\mu\nu}$ vanishes. As a result, equations \eqref{eq1}-\eqref{eq3} are trivially satisfied and equation \eqref{eq4} reduces to the Einstein's equation in the presence of the cosmological constant which has a de Sitter solution with the Hubble parameter of the form
\begin{align}\label{ds}
H=\sqrt{\f{\Lambda}{3}}.
\end{align}
Note that in an isotropic universe, the non-local Proca part of the action does not contribute to the evolution of the universe. 
It should be noted that in the local Proca theory, the equation of motion of the vector field gives $A_0=0$ and as a result the theory again adopt exact de Sitter solution. However in the non-local version of the theory, the gauge invariance implies that $A_0$ becomes arbitrary. So in the both form of the Proca action the vector field does not contribute the field equations in the FRW universe.
In the following, we will see that the non-local Proca term affects the perturbations around de Sitter space time. Also, in the next section, we will consider the anisotropic universe in which the non-local interaction term will affect the evolution of the universe.
\subsection{Cosmological perturbations}
Let us consider the perturbation of the action \eqref{ac1} on top of the de Sitter solution obtained in the previous section. The metric perturbation can be written as
\begin{align}\label{100}
ds^2&=-(1+2\varphi)\,dt^2+2a(S_i+\partial_i B)dx^i\, dt\nonumber\\&+a^2\big((1+2\psi)\delta_{ij}+\partial_i\partial_j E+\partial_{(i}F_{j)}+h_{ij}%
\big)dx^i dx^j,
\end{align}
where $h_{ij}$ is the traceless and transverse tensor mode with $h_{ii}=0=\partial_i h_{ij}$, $F_i$ and $S_i$ are transverse vector modes with $\partial_iF_i=0=\partial_iS_i$, and $\psi$, $\varphi$, $B$ and $E$ are four scalar modes. Note that the spacial indices  $(i,j=1,2,3)$ are raised and lowered with $\delta_{ij}$.

The vector field $A_\mu$ can be decomposed as
\begin{align}\label{101}
A_\mu=(A_0+\delta A_0,A^\perp_i+\partial_i \delta A),
\end{align}
where $\delta A_0$ and $\delta A$ are scalar modes and $A^\perp_i$ is the vector mode with $\partial_i A^\perp_i=0$. Note that the background value $A_0$ is an arbitrary function of time since it is not contribute to the background cosmology. So, in this section we will assume that $A_0$ is constant for the sake of simplicity. For the antisymmetric tensor fields $\xi_{\mu\nu}$ and $\lambda_{\mu\nu}$, the background values are zero and one has a decomposition
\begin{align}
\xi_{i0}&=\xi^\perp_i+\partial_i\xi,\nonumber\\
\xi_{ij}&=\epsilon_{ij}^{~~k}(\chi^\perp_k+\partial_k\chi),
\end{align}
and
\begin{align}
\lambda_{i0}&=\lambda^\perp_i+\partial_i\lambda,\nonumber\\
\lambda_{ij}&=\epsilon_{ij}^{~~k}(\rho^\perp_k+\partial_k\rho),
\end{align}
where $\xi^\perp_i$, $\chi^\perp_i$, $\lambda^\perp_i$ and $\rho^\perp_i$ are vector modes with vanishing divergence, and $\xi$, $\chi$, $\lambda$ and $\rho$ are scalar modes. Also, $\epsilon_{ij}^{~~k}$ is the 3-dimensional Levi-Civita symbol.  In summary, we have one tensor mode associated to the metric perturbation, 7 vector modes and 10 scalar modes. Now, let us define the gauge invariant quantities associated to above perturbation variables. For the metric perturbation, one can define two gauge invariant scalar perturbations of the form
\begin{align}
\Phi&=\varphi+\partial_t\left(aB-\f{a^2}{2}\partial_t E\right),\nonumber\\
\Psi&=\psi+H\left(aB-\f{a^2}{2}\partial_t E\right),
\end{align}
and one gauge invariant vector perturbation
\begin{align}
\Pi_i=S_i-\f12a\partial_t F_i.
\end{align}
Also, the tensor perturbation $h_{ij}$ is gauge invariant. For the vector field $A_\mu$, one can define two gauge invariant scalar perturbations of the form
\begin{align}
\mathcal{Y}&=\delta A_0+A_0\partial_t\left(aB-\f{a^2}{2}\partial_t E\right),\nonumber\\
\mathcal{Z}&=\delta A+A_0\left(aB-\f{a^2}{2}\partial_t E\right).
\end{align}
The vector mode $A^\perp_i$ is gauge invariant. Finally, one can easily check that     because of the zero background values for the antisymmetric tensor fields $\xi_{\mu\nu}$ and $\lambda_{\mu\nu}$, all vector and scalar modes associated to these fields are gauge invariant.

After expanding the action \eqref{ac1} up to second order in perturbation parameters around the de Sitter background \eqref{ds}, one can see that tensor, vector and scalar modes are decoupled from each other and as a result, in the following we will consider these modes independently.
\vspace{0.5cm}
\subsubsection{Tensor perturbations}
Expanding the action \eqref{ac1} up to second order in the tensor mode $h_{ij}$, one obtains
\begin{align}
S^{(2)}_{tensor}=\f12\sum_{+,\times}\int\, d^3k\,dt\, \kappa^2\,a^3 \left[\dot{h}^2_{ij}-\f{\vec{k}^2}{a^2}h^2_{ij}\right],
\end{align}
where we have performed Fourier transformation and sum over two helicity degrees of freedom.

The above result shows that the tensor mode in Non-local Proca theory is equivalent to the standard Einstein's theory with a cosmological constant. This is however not surprising since we do not have a source for the tensor mode from the non-local Proca interaction. As a result the speed of the propagation of tensor mode in this theory is equal to the speed of light, satisfying recent gravitational wave observations \cite{gw}.
\subsubsection{Vector perturbation}
Expanding the action \eqref{ac1} up to second order in gauge invariant perturbation vector modes and performing Fourier transformation, one obtains
\begin{widetext}
\begin{align}\label{vecact}
S^{(2)}_{vector}&=\int d^3kdt a\Bigg[\dot{\vec{A}}^{\perp2}-\f{\vec{k}^2}{a^2}\vec{A}^{\perp2}+\kappa^2 \vec{k}^2\vec{\Pi}^2+4\left(\f{\vec{k}^2}{a^2}-\f{4\Lambda}{3}\right)(\vec{\lambda}^\perp.\vec{\xi}^\perp-\vec{\rho}^\perp.\vec{\chi}^\perp)+\sqrt{\f{16\Lambda}{3}}(\vec{A}^\perp.\vec{\xi}^\perp+\dot{\vec{\xi}}^\perp.\vec{\lambda}^\perp-\dot{\vec{\chi}}^\perp.\vec{\rho}^\perp)\nonumber\\
&+4(\dot{\vec{A}}^\perp.\vec{\lambda}^\perp+\ddot{\vec{\xi}}^\perp.\vec{\lambda}^\perp-\ddot{\vec{\chi}}^\perp.\vec{\rho}^\perp)+4\alpha \vec{A}^\perp.\dot{\vec{\xi}}^\perp+\f{4}{a}(\vec{\rho}^\perp-\alpha\vec{\chi}^\perp).(\vec{k}\times\vec{A}^\perp)+\f{8H}{a^2}\Big(\vec{\rho}^\perp.(\vec{k}\times\vec{\xi}^\perp)+\vec{\lambda}^\perp.(\vec{k}\times\vec{\chi}^\perp)\Big)
\Bigg].
\end{align}
\end{widetext}
One can see from the above action that $\vec{\Pi}$, is non-dynamical with equation of motion $\vec{\Pi}=0$. We then substitute back this equation into the action. After integration by parts the kinetic part of the above action can be written as 
\begin{align}
S^{(2)}_{vector,kin}=\int d^3k\,dt\,a \left(\dot{\vec{\Sigma}}_{vec}^t\, \mathcal{K}_{vec}\,\dot{\vec{\Sigma}}_{vec}\right),
\end{align}
where $\mathcal{K}_{vec}$ is a $5\times5$ matrix and the matrix $\vec{\Sigma}_{vec}$ is defined as
\begin{align}
\vec{\Sigma}_{vec}=\left(\vec{A}^\perp ,\vec{\xi}^\perp ,\vec{\lambda}^\perp ,\vec{\chi}^\perp ,\vec{\rho}^{\perp}\right).
\end{align}
The eigenvalues of the kinetic matrix $\mathcal{K}_{vec}$ are
$(1,2,2,-2,-2)$ and the eigenvectors are $\vec{A}^\perp$  and
\begin{align}
&\vec{\mc{Y}}_1= \vec{\xi}^\perp+\vec{\lambda}^\perp,\quad \vec{\mc{Y}}_2=\vec{\xi}^\perp-\vec{\lambda}^\perp,\nonumber\\
&\vec{\mc{Y}}_3= \vec{\rho}^\perp+\vec{\chi}^\perp,\quad 
\vec{\mc{Y}}_4= \vec{\rho}^\perp-\vec{\chi}^\perp.\nonumber\\
\end{align}
In the subhorizon limit where $k\gg H=\sqrt{\Lambda/3}$, the action \eqref{vecact} reduces to
\begin{widetext}
\begin{align}\label{subvec}
S^{(2)}_{vector}\sim  \int d^3kdt a\Bigg[\dot{\vec{A}}^{\perp2}-\f{\vec{k}^2}{a^2}\vec{A}^{\perp2}+\dot{\vec{\mc{Y}}}_2^{\,2}+\dot{\vec{\mc{Y}}}_3^{\,2}-\f{\vec{k}^2}{a^2}\left(\vec{\mc{Y}}_2^{\,2}+\vec{\mc{Y}}_3^{\,2}\right)-
\bigg(\dot{\vec{\mc{Y}}}_1^{\,2}+\dot{\vec{\mc{Y}}}_4^{\,2}-\f{\vec{k}^2}{a^2}\left(\vec{\mc{Y}}_1^{\,2}+\vec{\mc{Y}}_4^{\,2}\right)\bigg)\Bigg].
\end{align}
\end{widetext}
One can see that in the subhorizon limit, there are 5 non-coupled dynamical transverse vector modes. Two of these vector fields i.e. $\vec{\mc{Y}}_1$ and $\vec{\mc{Y}}_4$ have wrong sign in the action \eqref{subvec}. However, as mentioned before the tensor field $\xi_{\mu\nu}$ is an auxiliary field which has been introduced from the localization of the action \eqref{ac}, and there are no quanta associated with it. 
The non-local action \eqref{ac} only contains the vector field $A_{\mu}$ and to solve the field equations we need the initial conditions only on this field. However the localized action \eqref{ac1} contains an auxiliary field $\xi_{\mu\nu}$  which is the specific solution of the equation $\Box\xi_{\alpha \beta}=F_{\alpha\beta}$, i.e. with fixed initial conditions $\xi_{\mu\nu}=0$ when $F_{\mu\nu}=0$. For example, with these conditions there are no arbitrary plane waves associated to $\xi_{\mu\nu}$ in flat space  \cite{mag}.
 
\hspace{0.2cm}

\subsubsection{Scalar perturbations}
Let us now consider the scalar perturbation of the theory \eqref{ac1}. As is obtained from the previous section, the theory has 8 gauge invariant scalar degrees of freedom. After Fourier transforming the second order action in scalar modes, one obtains
\begin{widetext}
\begin{align}\label{sca1}
S^{(2)}_{scalar}=\int\, d^3k\,dt\,\mathit{a} \,k^2\,\Bigg[&4  \sqrt{\f{\Lambda}{3} }\left( \lambda\dot\xi-\rho  \dot\chi\right) +4  \left( \lambda\ddot\xi -\rho  \ddot\chi\right) -2 (2 \alpha  \xi -2 \lambda +\mathcal{Y}) \dot{\mathcal{Z}}+ \dot{\mathcal{Z}{}}^2+ \f{4 \mathit{a}^2  \kappa ^2}{k^2} \left(2 \sqrt{3\Lambda } \Phi 
\dot\Psi -3 \dot\Psi {}^2- \Lambda  \Phi
^2\right)\nonumber\\
&+4 \left(\frac{ k^2 }{\mathit{a^2}}-\f43 \Lambda\right)\left(\lambda\xi-\rho\chi\right)+4  \kappa ^2  \left(2\Phi  \Psi + \Psi ^2\right)+ \left(\mathcal{Y}+4 \alpha   \xi  -4 \lambda  \right)\mathcal{Y}\Bigg].
\end{align}
\end{widetext}
One can see from the above action that  scalar perturbations $\Phi$, $\mc{Y}$, $\lambda$ and $\rho$ are non-dynamical, with equations of motion
\begin{align}
&\Lambda\Phi=\f{k^2}{a^2}\Psi+\sqrt{3\Lambda}\dot{\Psi},\label{phi}\\
&\mc{Y}=2\lambda+2\alpha\xi+\dot{\mc{Z}},\label{Y},
\end{align}
and
\begin{align}
2\lambda=\ddot{\xi}+\sqrt{\f{\Lambda}{3}}\dot{\xi}-\left(\f{k^2}{a^2}-4\Lambda-2\alpha\right)\xi.\label{lam}
\end{align}
Substituting above solutions  back to the action \eqref{sca1}, one can see that the scalar mode $\Psi$ becomes non-dynamical with equation of motion $\Psi=0$. This is similar to the Einstein-Hilbert theory since there is no non-minimal interactions between the curvature tensor and the fields $\xi_{\mu\nu}$ and $A_{\mu}$. However after substituting
 \eqref{lam} in the action \eqref{sca1}, we would have term $\ddot{\xi}^2$  which is well-known to suffer from Ostrogradski instability. To consider the dynamical modes in details, we  only substitute equations \eqref{phi} and \eqref{Y} in the action \eqref{sca1}. The kinetic part of the action then can be written as follow
\begin{align}
S^{(2)}_{scalar,kin}=\int d^3k\,dt\,a\, k^2\, \left(\dot{\vec{\Sigma}}_{scal}^t\, \mathcal{K}_{scal}\,\dot{\vec{\Sigma}}_{scal}\right),
\end{align}
where $\mathcal{K}_{scal}$ is a $4\times4$ matrix and the matrix $\vec{\Sigma}_{scal}$ is defined as
\begin{align}
\vec{\Sigma}_{scal}=\left(\xi,\lambda,\chi,\rho\right).
\end{align}
The kinetic matrix of the scalar modes $\mathcal{K}_{scal}$ has the eigenvalues $(-2,-2,2,2)$ which signals that two scalar modes are ghost. 

\begin{align}
&\mc{Y}_1= \xi+\lambda,\quad \mc{Y}_2=\xi-\lambda,\nonumber\\
&\mc{Y}_3= \rho+\chi,\quad 
\mc{Y}_4= \rho-\chi.
\end{align} 
Substituting above new scalar fields in the action \eqref{sca1} and considering the                                              subhorizon limit $k\gg H=\sqrt{\Lambda/3}$ we would have
\begin{align}
S^{(2)}_{scalar}&\sim\int\, d^3k\,dt\,a\, k^2\, \Bigg[\dot{\mc{Y}}_2^2+\dot{\mc{Y}}_3^2-\f{k^2}{a^2}\left(\mc{Y}_2^2+\mc{Y}_3^2\right)\nonumber\\&
-\dot{\mc{Y}}_1^2-\dot{\mc{Y}}_4^2+\f{k^2}{a^2}\left(\mc{Y}_1^2+\mc{Y}_4^2\right)\Bigg].
\end{align}
The above action shows that we have four non-coupled scalar modes in the subhorizon limit where two of them are healthy and the rest i.e. $\mc{Y}_1$ and $\mc{Y}_4$ have the wrong sign in the action.  This situation is similar to the vector perturbations. The scalar parts of the perturbed tensor field $\xi_{\mu\nu}$ does not have any quanta. As a result the number of degrees of freedom should be counted by the original theory.

\section{Anisotropic cosmology}\label{aniso}
Let us now consider the anisotropic cosmology of the non-local Proca electrodynamics.  Assuming that the universe can be described by the Bianchi type-I space-time of the form
\begin{align}
ds^2=-dt^2+a^2dx^2+b^2(dy^2+dz^2),
\end{align}
where $a(t)$ and $b(t)$ are the scale factors corresponding to spacial directions $x$, $y$ and $z$.
In order to compare the results with observations we will consider the matter fields minimally coupled to the non-local Proca field and metric tensor. We suppose that the energy momentum tensor is associated to the perfect fluid 
\begin{align}
T^\mu_\nu=dig(-\rho,p,p,p)
\end{align}
with equation of state $p=0$ representing dust dominated universe. This is plausible since radiation has a very small fraction of matter abundance at the late time.
Let us define the following quantities which are very useful in the analysis of anisotropic cosmology
\begin{align}
H_1&=\f{\dot{a}}{a},\quad H_2=\f{\dot{b}}{b},\quad \Delta H_i=H_i-H,\nonumber\\
3H&=\sum_i‌ H_i=H_1+2H_2,\nonumber\\
\quad 3A&=\sum_i\left(\f{\Delta H_i}{H}\right)^2,\quad
q=-1+\f{d}{dt}\left(\f{1}{H}\right).
\end{align}
Here, $H$ is the mean Hubble parameter, $A$ is the anisotropic parameter and $q$ is the deceleration parameter.
Since the matter field does not have non-minimal interaction with the vector field and the metric tensor, so it is conserved. The conservation of the matter field leads to 
\begin{align}
\dot{\rho}+3 H\rho=0.
\end{align}
In order to solve the set of equations \eqref{eq1}-\eqref{eq4}, let us perform some assumptions. From equation \eqref{eq1}, one obtains
\begin{align}
	\nabla^\beta(F_{\alpha\beta}-2\alpha\xi_{\alpha\beta}+2\lambda_{\alpha\beta})=0,
\end{align}
which has a solution
\begin{align}\label{sol1}
\lambda_{\alpha\beta}=\alpha\xi_{\alpha\beta}-\f12 F_{\alpha\beta},
\end{align}
where we have set the integration constant to zero. Note that $F_{\mu\nu}$ is the field strength of the vector field $A_{\mu}$. Inspiring from solution \eqref{sol1}, one can assume that $\xi_{\alpha\beta}$ and $\lambda_{\alpha\beta}$ are also the field strength of some vector fields $\xi_\mu$ and $\lambda_\mu$ respectively
\begin{align}
&\xi_{\alpha\beta}=\nabla_\alpha\xi_\beta-\nabla_\beta\xi_\alpha,\nonumber\\&\lambda_{\alpha\beta}=\nabla_\alpha\lambda_\beta-\nabla_\beta\lambda_\alpha.
\end{align}
Substituting back to \eqref{sol1}, one obtains
\begin{align}
\lambda_\mu=\alpha\,\xi_\mu-\f12A_\mu.
\end{align}
Now, let us assume that the vectors $A_\mu$ and $\xi_\mu$ can be decomposed as
\begin{align}
A_{\mu}&=\kappa\left(0,\int a \, U_1  dt,\int b \, U_2dt,\int b \, U_2 dt\right),\nonumber\\
\xi_{\mu}&
=\f{\kappa}{H_0^2}\left(0, \int  a \,X_1 dt, \int b \, X_2dt, \int b \, X_2 dt\right),
\end{align}
where $H_0$ is the current Hubble parameter and $U_i$ and $X_i$ with $i=1,2$ are some arbitrary dimensionless functions of time.
The time component of the above vectors are set to zero, because they do not contribute to the field equations. Also, the complicated form of the above assumption will help us to write the field equations in a more compact form.

Now, let us define the following dimensionless parameters 
\begin{align}
\tau&=H_0 t,\quad H=H_0 h,\quad \alpha=\beta H_0^2,\nonumber\\\Lambda&=3 H_0^2\lambda, \quad \bar{\rho}=\f{\rho}{3\kappa^2 H_0^2}.
\end{align}
With the above assumptions, equations of motion  \eqref{eq1}-\eqref{eq4} reduces to
\begin{widetext}
\begin{align}\label{c1}
h^2&\left[3(\tilde{A}^2-1)+2(\tilde{A}^2+1)S_1 X_1+2(2-2\tilde{A}+5\tilde{A}^2)S_2 X_2\right]+3 \lambda+3\bar\rho\nonumber\\&\hspace{4cm}+\sum_{j=1}^{2}\bigg[j\left(S_j^\prime X_j^\prime -X_j S_j^{\prime\prime}+S_j X_j^{\prime\prime}-\frac{1}{4}U_j^2\right)\bigg]=0,
\end{align}
\begin{align}\label{c2}
h^\prime&\left[-2(\tilde{A}+1)+4(1-2\tilde{A})S_2 X_2\right]
-h^2\left[(\tilde{A}+1)^2(3+2 S_1 X_1)+2(-4+10\tilde{A}+5\tilde{A}^2)S_2 X_2\right]\nonumber\\&-h\left[2(1+4S_2X_2)\tilde{A}^\prime+3X_1^2\left(\f{S_1}{X_1}\right)^\prime+4(2\tilde{A}-1)\left(S_2X_2\right)^\prime\right]
+3\lambda+2U_2(S_2+\beta X_2)+S_1 X_1^{\prime\prime}-X_1 S_1^{\prime\prime}\nonumber\\&\qquad-\sum_{j=1}^{2}\bigg[j\left(S_j^\prime X_j^\prime +\frac{1}{4}U_j^2\right)\bigg]=0,
\end{align}
\begin{align}\label{c3}
h^2&\left[2Z_1+2S_1X_1-3+\tilde{A}(1-\tilde{A})(2Z_1+8S_2X_2+3)\right]+h^\prime\left[2Z_1-2+\tilde{A}(2Z_1+1)\right]+h\bigg[\f12\tilde{A}^\prime\left(1+2Z_1\right)+2(1+\tilde{A})Z_1^\prime\nonumber\\&
-3X_2^2\left(\f{S_2}{X_2}\right)^\prime\bigg]+3\lambda-\f14 U_1^2+2S_2 U_2
+U_1(S_1+\beta X_1)-S_1^\prime X_1^\prime-2S_2^\prime X_2^\prime-X_2 S_2^{\prime\prime}+S_2 X_2^{\prime\prime}=0,
\end{align}
\begin{align}\label{c4}
2 &(\tilde{A}-2)\left[h^\prime Z_2+3h^2\left( S_2 X_1-S_1 X_2+ \tilde{A}Z_2\right)\right] +2h\left[\tilde{A}^\prime Z_2 +\tilde{A}\left(4Z_2^\prime-6(S_1 X_2^\prime+X_1 S_2^\prime)\right)+Z_2-6(X_1 S_2^\prime+X_2 S_1^\prime)\right]\nonumber\\&
-4\beta U_2 X_1+U_1(2U_2-4\beta X_2)+2(S_1X_2^{\prime\prime}+S_2 X_1^{\prime\prime}-X_2 S_1^{\prime\prime}-X_1S_2^{\prime\prime})=0,
\end{align}
\begin{align}\label{c5}
-2&(\tilde{A}+1)\left[3h^2+2h^\prime\right]S_2 X_2+h\left[5X_2S_2^\prime+S_2\left(2 \tilde{A}^\prime X_2-X_2^\prime\right)+2\tilde{A}\left(S_2X_2\right)^\prime\right]+\f12 U_2\left(U_2-4\beta X_2\right)\nonumber\\&-X_2 S_2^{\prime\prime}+S_2X_2^{\prime\prime}=0,
\end{align}
\end{widetext}
\begin{align}\label{c6}
U_1^{\prime\prime}+3hU_1^\prime+U_1\left(4\beta-2h^2(1+\tilde{A})^2\right)=0,
\end{align}		
and
\begin{align}\label{c7}
U_2^{\prime\prime}+3hU_2^\prime+U_2\left(4\beta-h^2(2-2\tilde{A}+5\tilde{A}^2)\right)=0,
\end{align}

where prime represents derivative with respect to $\tau$ and we have defined 
$\tilde{A}=\sqrt{A/2}$ and 
\begin{align}
S_j&=\beta X_j-\f12 U_j,\nonumber\\
Z_1&=S_1 X_1+S_2 X_2,\nonumber\\
Z_2&=S_1 X_2+S_2 X_1.\nonumber
\end{align}
\begin{figure*}[htbp]
	\centering
	\includegraphics[width=8.4cm]{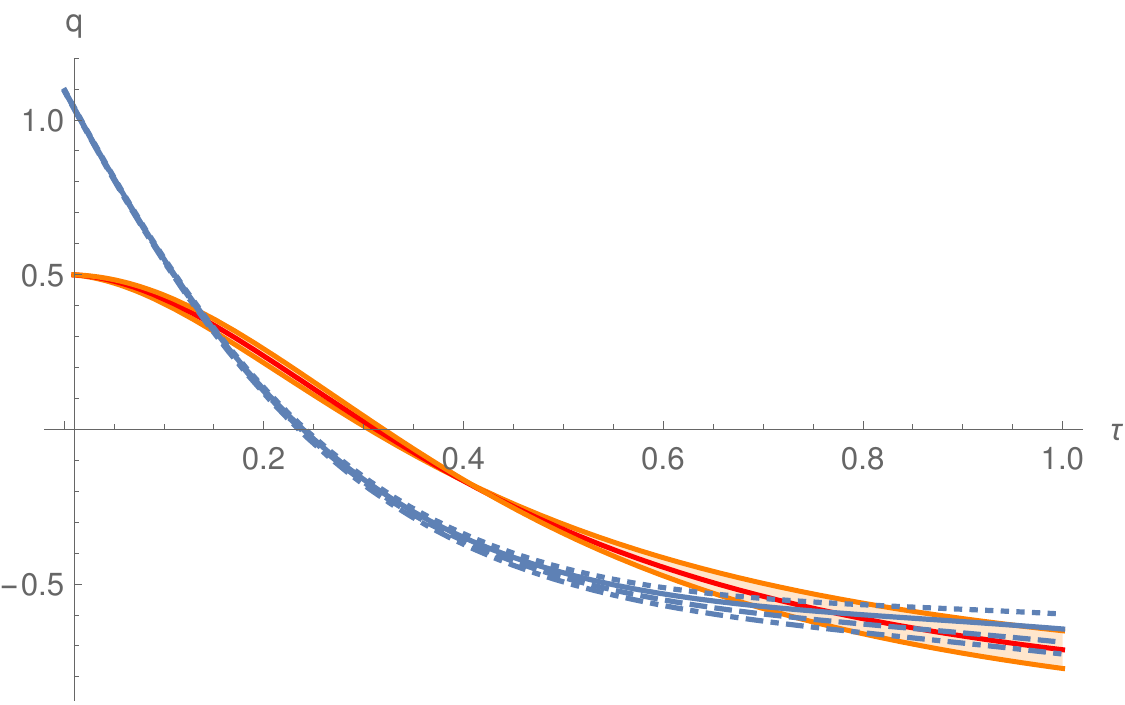}
	\includegraphics[width=8cm]{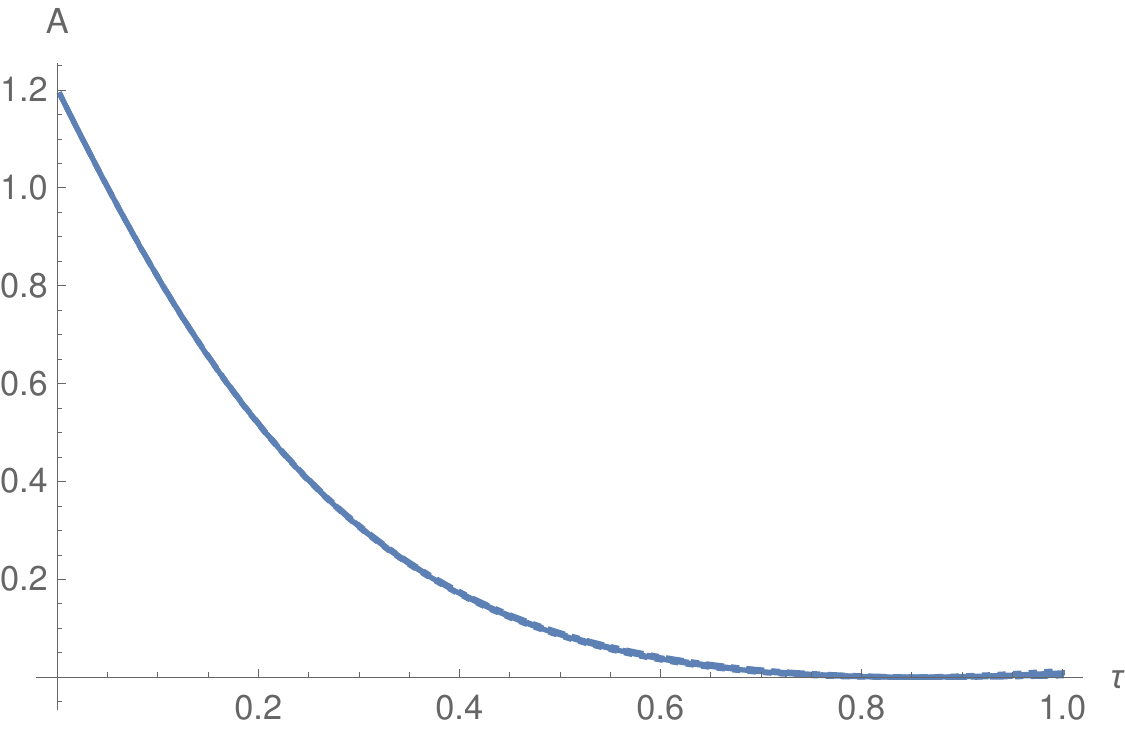}
	\caption{Variation of the deceleration parameter $q$ (left panel) and the anisotropy parameter $A$ (right panel) as a function of $\tau$ for $\lambda=0.805$ and different values of the $\beta$: $\beta=0.11$ (solid curve),   $\beta=0.10$  (dotted curve),  $\beta=0.12$ (dashed curve) and $\beta=0.13$ (dot-dashed curve), respectively. The solid shadowed orange curve correpsond to the $\Lambda$CDM model \cite{Planck}.}\label{fig1}
\end{figure*}
In figures \ref{fig1}, we have depicted the deceleration parameter $q$ and anisotropy parameter $A$ in terms of the dimensionless time parameter $\tau$ for  $\lambda=0.805$ and four different values of $\beta=0.10,\,0.11,\,0.12$ and $0.13$ respectively. Also, in figures \ref{fig3}, we have plotted these parameters for $\beta=0.11$ and four different value of $\lambda=0.6,\, 0.7,\, 0.8$ and $0.9$ respectively. One should note that the current value of the Cosmological constant parameter $\lambda=\Omega_{\Lambda,0}=0.68$ \cite{Planck}.
\begin{figure*}[htbp]
	\centering
	\includegraphics[width=8.4cm]{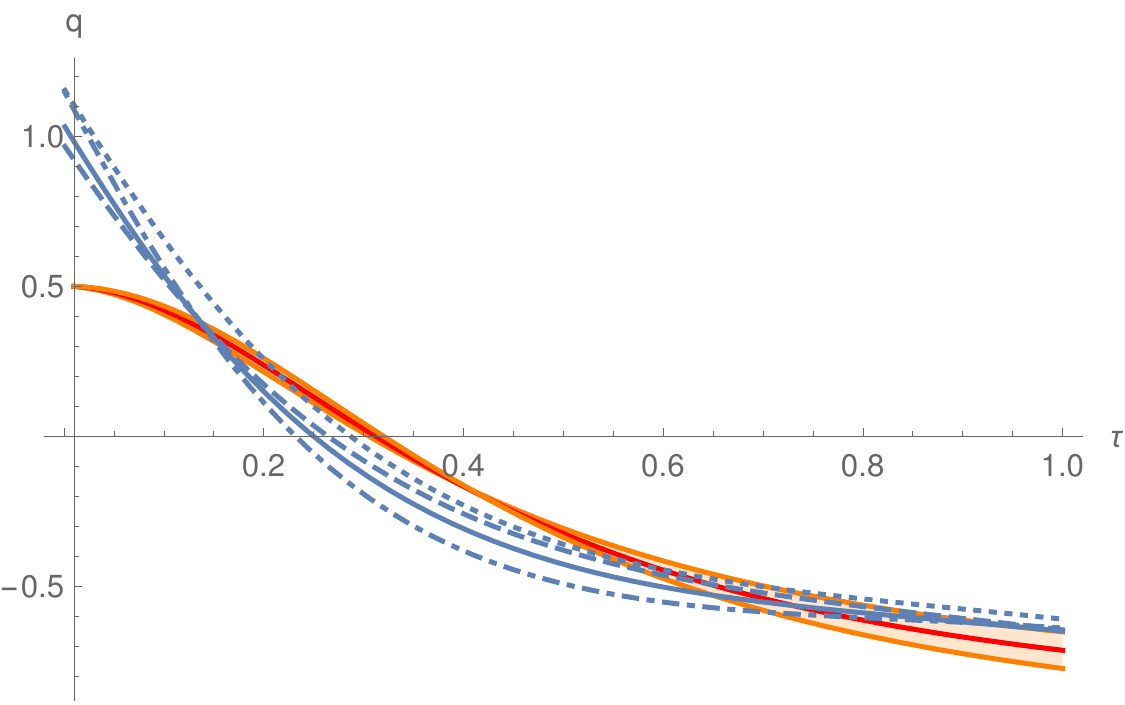}
		\includegraphics[width=8cm]{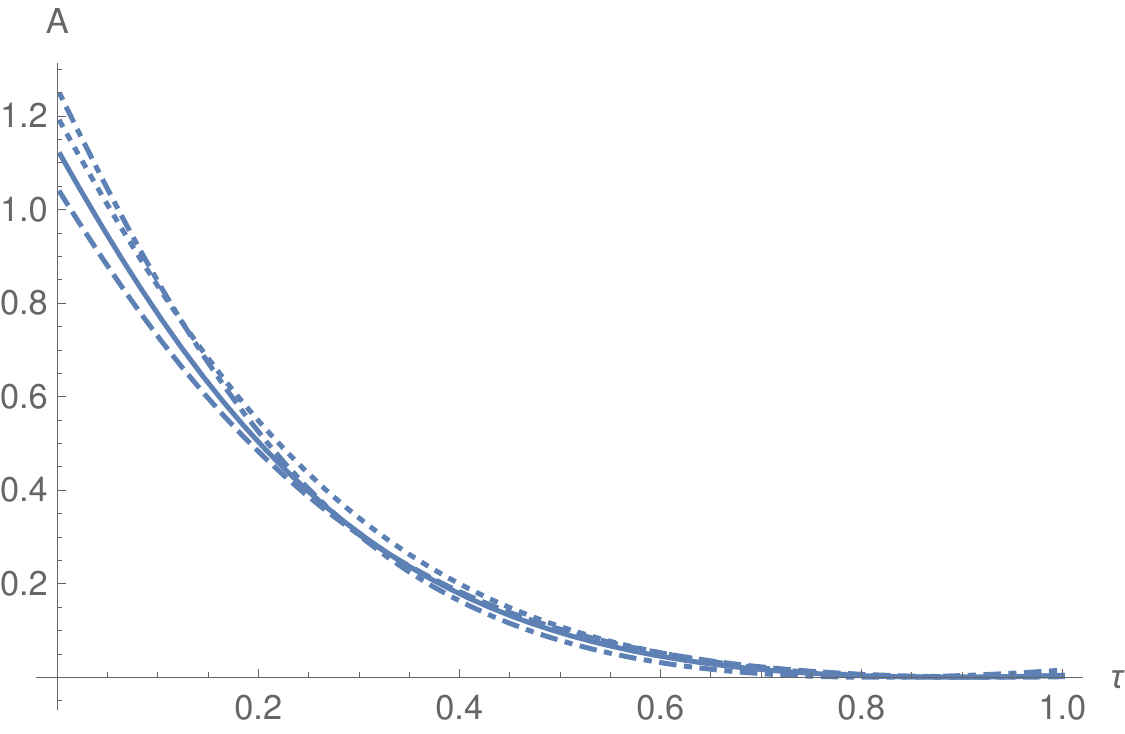}
	\caption{Variation of the deceleration parameter $q$ (left panel) and the anisotropy parameter $A$ (right panel) as a function of $\tau$ for $\beta=0.11$ and  different values of the parameter $\lambda$: $\lambda=0.8$ (solid curve),   $\lambda=0.6$ (dashed curve), $\lambda=0.9$ (dot-dashed curve) and $\lambda=0.7$ (dotted curve)  respectively.  The solid shadowed orange curve correpsond to the $\Lambda$CDM model \cite{Planck}.}\label{fig3}
\end{figure*}

The figures show that the late time behavior of the theory is consistent with the observed dynamics of the late time universe; the dynamic of universe begins from a highly anisotropic state and at the late time the anisotropy parameter goes to zero which dictates that universe becomes isotropic. Also, the study of the deceleration parameter, shows that the universe begins from decelerating phase which enters to the accelerated expansion phase at late times. We have also depicted the $\Lambda$CDM curve in the left panel of figures \ref{fig1} and \ref{fig3}. One can see from the figures that compare to the $\Lambda$CDM model, the non-local Proca theory predicts more decelerated universe at early times, and enters to the accelerating phase earlier that $\Lambda$CDM model.

\section{Conclusions}
In this paper we have considered the non-local Proca electrodynamics which has $U(1)$ symmetry. This theory in flat space time  reduces to the co-called Proca action which breaks  $U(1)$ symmetry describes a massive spin-1 field. However, in curved space-times, the action differs from Einstein-Proca theory. In fact space-times, one can transform the action to a local theory that respects  $U(1)$ symmetry with an additional Lagrange multiplier. 

The theory admits a de Sitter solution on the FRW background trivially, since the vector field can not  contribute to an isotropic and homogeneous universe. Cosmological perturbations on top of this de Sitter solution, reveals that the tensor mode is identical to $\Lambda$CDM model. This shows that the propagation speed of the gravitational waves in this theory is equal to the speed of light, satisfying recent gravitational wave observations \cite{gw}.

 The vector mode of the theory has apparently five transverse vector degrees of freedom. By diagonalizing the kinetic part of the action one can see that two vector modes have the wrong sign in the action. However in the subhorizon limit all vector modes decouple from each other. The boundary condition on the auxiliary field $\xi_{\mu\nu}$ eliminates the quanta associated with this field. The behavior of the scalar modes is similar to the vector modes. In this case there are four scalar modes which by diagonalizing the kinetic matrix we found that two of them have wrong sign. However these modes decouple from each other in the subhorizon limit.

As we have discussed above, the non-local Proca action does not contribute to the homogeneous and isotropic FRW universe. So, we have investigated the dynamics of the universe in anisotropic Bianchi-I universe. We have seen that the universe starts from a highly anisotropic and decelerating phase and at late times, the universe becomes isotropic and accelerating. As a result the model satisfies observational data on the late time behavior of the universe.

\appendix
\section{Uniqueness of de Sitter solution}\label{app}
Let us consider the flat FRW space-time in spherical coordinate as
\begin{align}
ds^2=-dt^2+a^2\left(dr^2+r^2d\Omega^2\right).
\end{align}
The form of the vector field that respect to the isotropic and homogeneity of the space-time is 
\begin{align}
A_{\mu}=\left(A_0,A_r,0,0\right),
\end{align}
where $A_0$ and $A_r$ are only functions of time $t$.
The tensor fields $\xi_{\mu\nu}$ and $\lambda_{\mu\nu}$ are taken as
\begin{align}
\xi_{0i}&=\left(\xi_1,0,0\right),\quad \xi_{ij}=\epsilon_{ij}^{~~k}\mathcal{X}_k,\quad \mathcal{X}_k=\left(\xi_2,0,0\right),\nonumber\\
\lambda_{0i}&=\left(\lambda_1,0,0\right),\quad \lambda_{ij}=\epsilon_{ij}^{~~k}\mathcal{Y}_k,\quad \mathcal{Y}_k=\left(\lambda_2,0,0\right),
\end{align}
where $\xi_i=\xi_i(t)$ and  $\lambda_i=\lambda_i(t)$ with $i=1,2$. 
The third component of the vector field equation \eqref{eq1} leads to
\begin{align}
\alpha\,\xi_2=-\lambda_{2}.
\end{align}
The (03) component of equation \eqref{eq2} then leads to $\lambda_{2}=0$. So by the use of the above relation we also have  $\xi_{2}=0$.
The (01) components of equations \eqref{eq2} and \eqref{eq3} yield
$\xi_1=0$ and $\lambda_1=0$ respectively. Substituting the above results in the (0) component of the equation \eqref{eq1}, one can obtain $\dot{A_r}=0$. Since the components of the vector field $A_\mu$  appears in the field equations only through the strength tensor $F_{\mu\nu}$, the $A_0$ and $A_r$ components with the condition $\dot{A_r}=0$ does not have any contribution in field equations.

\end{document}